\documentclass[conference]{IEEEtran}
\IEEEoverridecommandlockouts

\usepackage{cite}
\usepackage{amsmath,amssymb,amsfonts}
\usepackage{graphicx}
\usepackage{textcomp}
\usepackage{xcolor}
\usepackage{subcaption}
\usepackage{mathtools}
\usepackage{algorithm}
\usepackage{algpseudocode}
\usepackage{multirow}
\usepackage[hidelinks]{hyperref}

\definecolor{tueScharlaken}{HTML}{C81919}

\def\BibTeX{{\rm B\kern-.05em{\sc i\kern-.025em b}\kern-.08em
    T\kern-.1667em\lower.7ex\hbox{E}\kern-.125emX}}
\begin{document}

\setlength{\textfloatsep}{1mm} 

\title{Target Speaker Selection for Neural Network Beamforming in Multi-Speaker Scenarios\\
\thanks{This work was supported by the Robust AI for SafE (radar) signal processing (RAISE) collaboration framework between Eindhoven University of Technology and NXP Semiconductors, including a Privaat-Publieke Samenwerkingen-toeslag (PPS) supplement from the Dutch Ministry of Economic Affairs and Climate Policy.}
}

\makeatletter
\newcommand{\linebreakand}{%
  \end{@IEEEauthorhalign}
  \hfill\mbox{}\par
  \mbox{}\hfill\begin{@IEEEauthorhalign}
}
\makeatother

\newlength{\extrawidth}
\settowidth{\extrawidth}{\textit{iiiii}}

\newlength{\extrawidthb}
\settowidth{\extrawidthb}{\textit{iiIIIIIiiiiii}}


\author{\IEEEauthorblockN{Luan Vinícius Fiorio}
\IEEEauthorblockA{\textit{Department of Electrical Engineering} \\
\textit{Eindhoven University of Technology}\\
Eindhoven, The Netherlands \\
l.v.fiorio@tue.nl}
\and
\IEEEauthorblockN{Bruno Defraene}
\IEEEauthorblockA{\hspace{\extrawidth}\textit{NXP Semiconductors}\hspace{\extrawidth}\\
Leuven, Belgium \\
bruno.defraene@nxp.com}
\and
\IEEEauthorblockN{Johan David}
\IEEEauthorblockA{\hspace{\extrawidth}\textit{NXP Semiconductors}\hspace{\extrawidth}\\
Leuven, Belgium \\
j.david@nxp.com}
\and
\IEEEauthorblockN{Alex Young}
\IEEEauthorblockA{\hspace{\extrawidth}\textit{NXP Semiconductors}\hspace{\extrawidth}\\
Eindhoven, The Netherlands \\
alex.young@nxp.com}
\and
\IEEEauthorblockN{Frans Widdershoven}
\IEEEauthorblockA{\hspace{\extrawidthb}\textit{NXP Semiconductors}\hspace{\extrawidthb}\\
Eindhoven, The Netherlands \\
frans.widdershoven@nxp.com}
\and
\IEEEauthorblockN{Wim van Houtum}
\IEEEauthorblockA{\hspace{\extrawidthb}\textit{NXP Semiconductors}\hspace{\extrawidthb}\\
Eindhoven, The Netherlands \\
wim.van.houtum@nxp.com}
\and
\IEEEauthorblockN{Ronald M. Aarts}
\IEEEauthorblockA{\textit{Department of Electrical Engineering} \\
\textit{Eindhoven University of Technology}\\
Eindhoven, The Netherlands \\
R.M.Aarts@tue.nl}
}

\maketitle

\begin{abstract}
We propose a speaker selection mechanism (SSM) for the training of an end-to-end beamforming neural network, based on recent findings that a listener usually looks to the target speaker with a certain undershot angle. The mechanism allows the neural network model to learn toward which speaker to focus, during training, in a multi-speaker scenario, based on the position of listener and speakers. However, only audio information is necessary during inference. We perform acoustic simulations demonstrating the feasibility and performance when the SSM is employed in training. The results show significant increase in speech intelligibility, quality, and distortion metrics when compared to the minimum variance distortionless filter and the same neural network model trained without SSM. The success of the proposed method is a significant step forward toward the solution of the cocktail party problem.
\end{abstract}


\section{Introduction}

``How do we recognize what one person is saying when others are speaking at the same time?'' \cite[p.~117]{bronkhorst2000cocktail}. This simple question formulates the \emph{cocktail party problem}, which refers to the ability of the human hearing to separate voices that are mixed, in frequency and time. While such an ability is present in normal hearing, hearing impaired listeners might face difficulty in segregating auditory streams \cite{bee2008cocktail}.

Hearing impaired listeners frequently rely on hearing aids, sound-amplifying devices which employ beamforming strategies. Such devices usually beamform in front of the listener, while recent findings show that the listener's head has a tendency to undershot the target speaker's position \cite{lu2021investigating}. Beamforming algorithms help improving speech intelligibility and sound quality \cite{kidd2015benefits}, however, in reverberant multi-speaker scenarios, the performance of algorithms such as the minimum variance distortionless response (MVDR) filter is reduced \cite{cauchi2015combination}. More recently, audio beamforming was developed using neural networks (NN), end-to-end \cite{chen2022multichannel} or estimating signals feeding into a beamforming filter \cite{zhang2021adl}. Such approaches usually do not take multi-speaker scenarios into account, limiting its application, or employ additional sensors (e.g., cameras) for guiding the beam, which can be prohibitive for most hearing aid devices.

Inspired from the findings of \cite{lu2021investigating} regarding the presence of an undershot angle between listener's head and speaker direction, we propose a speaker selection mechanism for the training of beamforming neural networks. The mechanism teaches the model to focus on the target speaker based on the smallest undershot angle, requiring only audio information during inference. Through acoustic simulations, we show that a neural network trained with the mechanism is able to outperform the baseline model, trained without it, and the MVDR filter \cite{souden2010mvdr}. We also show that the proposed algorithm is robust to changes in number and position of speakers, a significant progress toward solving the cocktail party problem.

\section{Preliminaries}
The problem we tackle consists of multi-microphone audio beamforming in a multi-speaker scenario, where the microphones are positioned as of simulating hearing aid devices wore by a listener. $N$ speakers and a listener are randomly positioned in a reverberant room. The listener can look toward one of the speakers directly, or with an \emph{undershot} azimuth angle, as pointed out in \cite{lu2021investigating}. For generality, we also consider the overshot (though we prioritize the term undershot for readability) when the listener looks further than the desired speaker angle. Our objective is to extract the clean reverberant speech of the desired speaker only using audio information. 

An example can be seen in Fig.~\ref{fig:examplescenario}, where in a reverberant room, a listener $L$ looks toward a speaker $S_2$ with an undershot angle $\theta_u$. In this case, $S_2$ is the desired speaker while $S_1$ is undesired. The undershot angle can be described in terms of the listener's head center axis angle $\theta_h$ and the angle of the desired speaker $\theta_{S_2}$ ($\theta_{S_1}$ for the undersired speaker), in relation to the listener's x-axis, as $|\theta_u| = |\theta_{h} - \theta_{S_2}|$. In this example, the objective would be to extract the reverberant speech of speaker $S_2$ as received by a reference microphone.

The output $y_m(t)$ of each microphone $m$ is defined by the speech fragment $s_n(t)$ of speaker $n$, convolutioned ($\ast$) with a room impulse response (RIR) $g_{nm}(t)$ from speaker $n$ to microphone $m$, summed for all speakers. This is described as
\begin{equation}
    y_m(t) = \sum_{n} s_{n}(t) \ast g_{n,m}(t).
\label{eq:conv}    
\end{equation}
Our objective is to extract the desired (subscript $d$) reverberant speech at the reference microphone $s_{n=d}(t) \ast g_{n=d,m=ref.}(t)$, solely given microphone outputs $y_m(t), \ \forall \ m \in [1,\ ...,\ M]$. Additional noise is not considered in order to facilitate the demonstration of the proposed method. 

Additionally, the system we use in this work operates in the time-frequency domain, where $Y_m(t,f)$ and $S_n(t,f)$ are, respectively, the short-term Fourier transform (STFT) of the microphone outputs and the reverberant speech signal $s_n(t)$ captured by a reference microphone.

\begin{figure}[!t]
    \centering
    \includegraphics[width=0.85\linewidth]{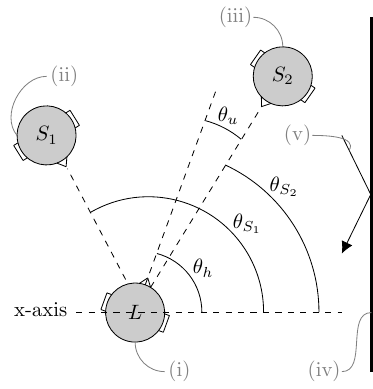}
    \caption{Example scenario of the considered problem. The indication arrows point out to: (i) listener $L$; (ii) speaker $S_1$; (iii) speaker $S_2$; (iv) wall; and (v) reverberation.}
    \label{fig:examplescenario}
\end{figure}

\begin{figure*}
    \centering
    \includegraphics[width=0.85\linewidth]{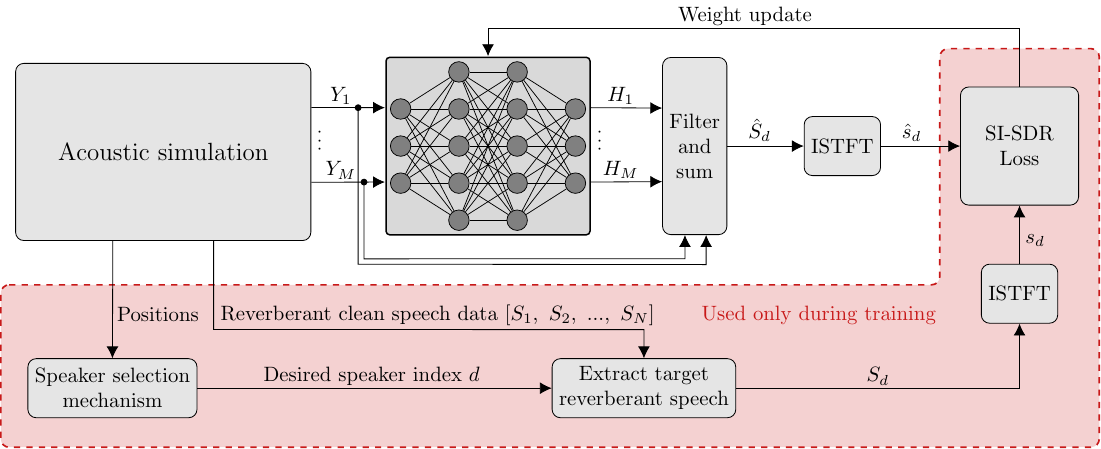}
    \caption{End-to-end beamforming neural network training system employing speaker selection mechanism.}
    \label{fig:system}
    \vspace{-5mm}
\end{figure*}

\begin{algorithm}
\caption{Speaker selection mechanism for two speakers}
\begin{algorithmic}[1]
\Procedure{SSM}{positions, $|\theta_{u}^{\max}|$}
    \State \textbf{Input:} 2-dimensional position of listener $[a_L, b_L]$ and speakers 
    $\Bigl[\bigl[a_{S_1}, b_{S_1}\bigr],\ \bigl[a_{S_2}, b_{S_2}\bigr]\Bigr]$ of an audio utterance
    \State \textbf{Parameter:} Maximum undershot angle $|\theta_{u}^{\max}|$
    \State \textbf{Output:} Index of desired speaker
    
    \State \textbf{Calculate the speakers' angles relative to the listener's x-axis:}
    \[
    \theta_{S_1} = \operatorname{atan2}(b_{S_1} - b_L,\; a_{S_1} - a_L)
    \]
    \[
    \theta_{S_2} = \operatorname{atan2}(b_{S_2} - b_L,\; a_{S_2} - a_L)
    \]
    
    \State \textbf{Determine the admissible range for $\theta_h$:} \\
    Ensure that the listener’s head angle is always closer than $|\theta_{u}^{\max}|$ from both speakers:
    \[
    \theta_{h}^{\min} = \min\{\theta_{S_1}, \theta_{S_2}\} - |\theta_{u}^{\max}|
    \]
    \[
    \theta_{h}^{\max} = \max\{\theta_{S_1}, \theta_{S_2}\} + |\theta_{u}^{\max}|
    \]
    
    \State \textbf{Sample the listener's head angle:}
    \[
    \theta_h \sim \operatorname{Uniform}(\theta_{h}^{\min},\; \theta_{h}^{\max})
    \]
    
    \State \textbf{Calculate the undershot angles for each speaker:}
    \[
    |\theta_{u1}| = |\theta_h - \theta_{S_1}|
    \]
    \[
    |\theta_{u2}| = |\theta_h - \theta_{S_2}|
    \]
    
    \State \textbf{Select the speaker:} 
    \If{$|\theta_{u1}| < |\theta_{u2}|$}
        \State \textbf{Return} 1 \Comment{Speaker 1 is desired}
    \Else
        \State \textbf{Return} 2 \Comment{Speaker 2 is desired}
    \EndIf
\EndProcedure
\end{algorithmic}
\label{alg:speakerselection}
\end{algorithm}

    
    
    

\section{Speaker selection mechanism}
\label{sec:mechanism}
We propose a speaker selection mechanism (SSM) for allowing a neural network to learn which speaker is desired, and beamform toward it. This approach can be applied to a NN that is: estimating the steering vector of a classical beamforming algorithm, like the MVDR filter \cite{zhang2021adl}; estimating the position of the target speaker \cite{grinstein2024neuralsrp}; estimating a time-frequency filter, which can be applied to the microphones' outputs via filter-and-sum operation \cite{chen2022multichannel}; among other uses. In this work, we choose to validate the proposed mechanism with an end-to-end neural network that estimates a multi-channel time-frequency beamforming filter.

Our method is inspired on the findings of \cite{lu2021investigating}, where it was noted that even though the eyes of the listener may follow precisely the speaker of interest, the head usually undershots the desired speaker. This is of crucial importance for applications where no visual or auxiliary cues are available, e.g., the case of most hearing aid devices. 

Nevertheless, we assume that we have access to all microphones' outputs in the array, and that the position of listener and speakers is known during training. The SSM works as follows. For each training utterance, we calculate the absolute value of the undershot angles for all speakers. We then identify the speaker that results in the smallest absolute undershot angle and set it as desired. Moreover, the desired speaker is used as a target for calculating the loss, during training, for that specific utterance. The target speaker in the loss function changes dynamically according to the smallest undershot angle. We consider the criteria of smallest undershot angle for changing desired speaker, but the movement of the head could be more explored, being out of scope for this paper. Alg.~\ref{alg:speakerselection} details the speaker selection mechanism for an example situation of two speakers. Notice that, in inference mode, there is no need to provide any information regarding position. The NN trained with the proposed mechanism is able to beamform toward the desired speaker solely with audio information.

Differently from \cite{veluri2024look}, our approach does not require the listener to face the target at any moment, and only one neural network is used. Visual cues, e.g., as considered in \cite{zhang2021adl}, are not taken into account in our method since we assume that the neural network can obtain spatial information based only on audio features. Next, we describe the model and simulation framework for evaluating the proposed SSM.

\section{Model and simulation framework}

We evaluate the SSM with an end-to-end neural network beamforming system, as per Fig.~\ref{fig:system}. A simulation environment outputs multi-microphone recordings, which are preprocessed and fed into the NN model in the time-frequency domain. The output of the neural network consists of a complex multi-channel filter $H_m(t,f) ,\ \forall m \in [1,\ ...,\ M]$, applied to the microphone recordings with a filter-and-sum operation, as 
\begin{equation}
    \hat{S}_d(t,f) = \sum_{m} Y_{m}(t,f) \cdot H_m(t,f).
\label{eq:filterandsum}    
\end{equation}
The model description is given in the following.

\subsection{Audio beamforming model}
\label{ssec:model}
We consider a NN-based beamforming approach with filter-and-sum, similar to \cite{chen2022multichannel}, but we simplify the model by using only real-valued operations, with a real-imaginary split at the input, concatenating both in the frequency axis. Consequentially, the output is recombined as a complex filter. Further on reducing the model's complexity, the convolutions are defined only in the frequency axis, as we did not observe significant performance difference against kernels in both frequency and time axis. The NN model is depicted in Fig.~\ref{fig:model}.

The model is trained to maximize the scale-invariant signal-to-distortion ratio (SI-SDR) of filtered microphone outputs in relation to the desired speaker's reverberant speech at the reference microphone. Differently from scale-invariant signal-to-noise ratio used in \cite{chen2022multichannel}, we consider the SI-SDR since it is a lower bound to both SDR and SNR \cite{roux2018sdrhalfbakeddone}.

For comparison, we train the same model twice. First, trained with the SSM for speaker-aware beamforming. Second, without using the proposed mechanism, by always setting a random speaker as the desired target, creating a NN baseline for the task that we are aiming to solve -- beamforming on multi-speaker scenarios with undershot angles. To the best of our knowledge, this is the first study to propose a solution for this task that neither requires the listener to face the speaker at any time nor relies on additional sensors. We also compare it to an ideal MVDR, always beamforming in the target direction. The MVDR approach is equivalent to the optimal case for when MVDR parameters are calculated with NNs, e.g., \cite{zhang2021adl}.

\begin{figure*}
    \centering
    \includegraphics[width=0.75\linewidth]{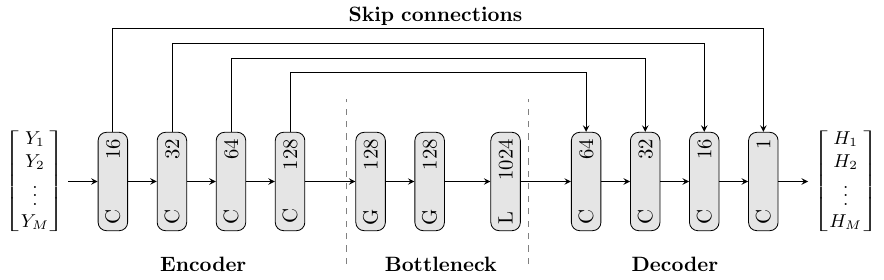}
    \caption{Schematic of the considered neural nerwork model. The number in each layer indicates output channels. The $\mathrm{C}$ encoder layers consist of Conv2D with BatchNorm2D and ReLU functions in all layers. An additional Tanh is applied to the encoder output to bound values, ensuring stable inputs for the recurrent layers. The $\mathrm{G}$ layers are gate recurrent units (GRUs), and $\mathrm{L}$ is a linear layer. The decoder layers are Conv2D.T, with BatchNorm2D and ReLU activation in all layers but the last, without normalization or activation. All $\mathrm{C}$ kernels are (8,1) with stride (2,1) and padding (3,0).}
    \label{fig:model}
    \vspace{-5mm}
\end{figure*}

\subsection{Acoustic simulation setup}
\label{ssec:sim}

For this evaluation, we simulate a reverberant room with four microphones and two speakers. First, the walls, floor and ceiling are defined, forming a rectangle-shaped room of size 5.15 x 3.75 x 2.65 m. Although the room size is fixed, the time it takes for sound pressure to reduce by 60 dB (T60) is defined over a variable range, assuring generality. The room impulse response (RIR) for each speaker in relation to each microphone is generated using gpuRIR \cite{diazguerra2021gpurir}. We set up the simulation as described in the following.

The microphones are omnidirectional and are positioned similarly as in a hearing aid device wore by a person. First, the position of the listener is (randomly) defined, and we assume a radius equal to 0.15 m, which is similar to the head breadth of an adult person. Two groups of two microphones are positioned in the east-most point and equivalently at the west-most point. The microphones within each group are split from each other by 0.50 cm, and positioned at the same height, with the front left microphone taken as reference.

Moreover, the speakers are randomly positioned following a few constraints. The first constraint is that the speakers cannot be closer than 1.00 m from the listener, and they cannot be closer than 1.00 m from each other. At the moment of positioning, the absolute angle difference of both speakers in relation to the listener must be of at least 45 degrees, avoiding that a speaker would be too close or behind the other speaker. Both listener and speaker positions are limited to be distant from any wall at least twice the head breadth value. The listener and speaker points are positioned with a height ranging from 1.50 and 1.95 m, similar to most adult humans' height. When all speakers and listener are positioned, the head angle of the listener is defined by randomly rotating the center of the two groups of microphones in the azimuth direction, but not exceeding a maximum undershot of 30 degrees. The maximum undershot constraint provides a better sense of reality, as a listener would not look too far from the desired speaker. 

Additionally, the signal-to-noise ratio (SNR), calculated with the mixed audio utterance (representing noise) against the desired-speaker-only utterance (representing signal) is randomly varied from -10 to 20 dB. The SNR is calculated considering the entire audio utterance, without filtering out silence periods. Note that speech traces are combined such that there is minimum silence period, but still sounding natural. Table~\ref{tab:parameters} summarizes the variable parameters in the simulation.

\begin{table}[!t]
    \centering
    \caption{Variable parameters and ranges for the acoustic simulation.}
    \begin{tabular}{c c c}
    \hline
        \textbf{Parameter} & \textbf{Min. value} & \textbf{Max. value} \\
        \hline
        T60 (s) & 0.20 & 1.00 \\
        SNR (dB) & -10.00 & 20.00 \\
        Listener/speaker height (m) & 1.50 & 1.95 \\
        Undershot angle (°) & -30 & +30 \\
    \hline
    \end{tabular}
    \label{tab:parameters}
\end{table}

\subsection{Data}
\label{ssec:data}
We use the LibriTTS dataset \cite{Zen2019Libritts} for the acoustic simulation. For each speaker, traces of speech are randomly selected and resampled to 16 kHz, combined until a duration of 10 seconds is reached, with a fade-in and fade-out of 0.05 to 0.20 seconds. Each speech trace is multiplied by a gain, randomly defined from -3 to 3 dB. Both speech utterances are adjusted to avoid clipping when combined. Each utterance is then convolved with the RIR referent to that speaker and microphones, which are obtained as described in Section~\ref{ssec:sim}, according to \eqref{eq:conv}, resulting in the microphone outputs. 

Moreover, the STFT operation is applied to the microphone outputs for 256 samples, with a Hann window of size 256 and a hop of 128 samples. The STFTs used as input to the neural network are normalized by their mean and standard deviation. Real and imaginary parts are then concatenated in the frequency axis, forming the input to the neural network. For training, the `train-clean-360' subset of LibriTTS is used, with 360 hours of raw audio. The evaluation is performed on the `test-clean' set, with approximately 8.6 hours of data.

\section{Results and discussion}

We train the neural network model described in Section~\ref{ssec:model} with and without the SSM proposed in Section~\ref{sec:mechanism}, for $N=2$ speakers, according to the acoustic parameters defined in Section~\ref{ssec:sim}, with the data mentioned in Section~\ref{ssec:data}. We also consider the (ideal) MVDR filter as a baseline, implemented as proposed in \cite{souden2010mvdr}. At every utterance, the MVDR filter is calculated based on the reference microphone with access to all separate (reverberant) signals, i.e., always beamforming in the target speaker direction. In Table~\ref{tab:results}, we show the average values over the `test-clean' set of LibriTTS of short-time objective intelligibility (STOI), perceptual evaluation of speech quality (PESQ), and SI-SDR, for the mixed audio at the reference microphone and the filtered signals. 

\begin{table*}[!t]
    \centering
    \caption{Average STOI, PESQ, and SI-SDR over the `test-clean' LibriTTS set for the mixed audio and the NN-filtered speech, trained with and without SSM for two speakers and evaluated for $N=2$ and $N=3$ speakers.}
    \begin{tabular}{c c c c c c c c c c c c c c c}
        \hline
        
        & & \multicolumn{3}{c}{-10 dB SNR} & \multicolumn{3}{c}{0 dB SNR} & \multicolumn{3}{c}{10 dB SNR} & \multicolumn{3}{c}{20 dB SNR} \\
        
        \textbf{N} & \textbf{Method} & \textbf{STOI} & \textbf{PESQ} & \textbf{SI-SDR} & \textbf{STOI} & \textbf{PESQ} & \textbf{SI-SDR} & \textbf{STOI} & \textbf{PESQ} & \textbf{SI-SDR} & \textbf{STOI} & \textbf{PESQ} & \textbf{SI-SDR} \\
        
        \hline        
        
        \multirow{4}{*}{2} & \textcolor{darkgray}{None (mixed)} & \textcolor{darkgray}{0.384} & \textcolor{darkgray}{1.237} & \textcolor{darkgray}{-9.970} & \textcolor{darkgray}{0.634} & \textcolor{darkgray}{1.535} & \textcolor{darkgray}{0.032} & \textcolor{darkgray}{0.849} & \textcolor{darkgray}{2.334} & \textcolor{darkgray}{10.031} & \textcolor{darkgray}{0.958} & \textcolor{darkgray}{3.431} & \textcolor{darkgray}{20.030} \\
                
        & MVDR filter & 0.447 & 1.322 & -6.445 & 0.682 & 1.740 & 2.043 & 0.838 & 2.430 & 5.109 & 0.868 & 2.707 & 1.866 \\
        
        & NN & 0.346 & 1.219 & -11.172 & 0.629 & 1.532 & -0.007 & 0.861 & 2.430 & 10.913 & \textcolor{tueScharlaken}{0.963} & 3.617 & \textcolor{tueScharlaken}{20.790} \\
        
        & NN + SSM training & \textcolor{tueScharlaken}{0.526} & \textcolor{tueScharlaken}{1.366} & \textcolor{tueScharlaken}{-1.608} & \textcolor{tueScharlaken}{0.736} & \textcolor{tueScharlaken}{1.851} & \textcolor{tueScharlaken}{5.012} & \textcolor{tueScharlaken}{0.891} & \textcolor{tueScharlaken}{2.790} & \textcolor{tueScharlaken}{12.541} & \textcolor{tueScharlaken}{0.963} & \textcolor{tueScharlaken}{3.746} & 20.227 \\
        \hline

        \multirow{4}{*}{3} & \textcolor{darkgray}{None (mixed)} & \textcolor{darkgray}{0.313} & \textcolor{darkgray}{1.237} & \textcolor{darkgray}{-9.990} & \textcolor{darkgray}{0.580} & \textcolor{darkgray}{1.465} & \textcolor{darkgray}{0.016} & \textcolor{darkgray}{0.828} & \textcolor{darkgray}{2.211} & \textcolor{darkgray}{10.017} & \textcolor{darkgray}{0.954} & \textcolor{darkgray}{3.368} & \textcolor{darkgray}{20.018} \\

        & MVDR filter & 0.371 & \textcolor{tueScharlaken}{1.285} & -7.314 & 0.634 & 1.626 & 1.786 & 0.827 & 2.327 & 5.690 & 0.872 & 2.714 & 2.238 \\
        
        & NN & 0.299 & 1.225 & -10.582 & 0.583 & 1.468 & 0.156 & 0.845 & 2.323 & 11.025 & 0.960 & 3.569 & \textcolor{tueScharlaken}{20.740} \\
        
        & NN + SSM training & \textcolor{tueScharlaken}{0.400} & 1.253 & \textcolor{tueScharlaken}{-6.311} & \textcolor{tueScharlaken}{0.669} & \textcolor{tueScharlaken}{1.652} & \textcolor{tueScharlaken}{3.361} & \textcolor{tueScharlaken}{0.874} & \textcolor{tueScharlaken}{2.621} & \textcolor{tueScharlaken}{12.222} & \textcolor{tueScharlaken}{0.961} & \textcolor{tueScharlaken}{3.703} & 20.273 \\
        \hline

        \hline        
    \end{tabular}
    \label{tab:results}
    \vspace{-5mm}
\end{table*}

We can see from Table~\ref{tab:results}, for $N=2$ speakers, that the use of the SSM in training can significantly increase the performance of the neural network-based beamforming model, for all considered SNRs. As expected, the proposed mechanism is able to teach the network which speaker is of interest at each utterance. When the model is trained without such information, a lower signal-to-noise ratio condition causes the performance to be drastically affected since the NN model ``confuses'' the choice of speaker. We can see that, as the SNR of the speech combination increases, the NN without SSM becomes able to separate desired from undesired speaker, indicating that it is focusing on the higher-amplitude signal, a major feature in the audio combination. However, even for higher SNR levels, the performance of the baseline NN is insufficient, as the model trained with SSM almost always forms an upper bound for the NN's performance. For the lower considered SNR levels (-10 and 0 dB), the NN without SSM cannot even surpass the metrics obtained with the mixed signal, received at the reference microphone. The NN trained with SSM, on the other hand, is able to extract the desired reverberant speech trace.

Moreover, the MVDR filter is outperformed by the NN with SSM training for almost all cases. The baseline NN provides a similar or better performance than the MVDR filter at higher SNRs. That is due to the MVDR formulation, which assumes an acoustic scene with anechoic conditions, while the NNs can learn to suppress the effects of reverberation. For higher SNR, the reverberation of the desired speaker has more energy, contaminating the direct path and deteriorating the MVDR performance, which can be noticed in terms of SI-SDR.

We also check the robustness of the proposed mechanism against changes in the environment by re-evaluating all methods for a different acoustic scenario. Now, we consider a more challenging case of $N=3$ speakers, with minimum distance between listener to speakers, and speakers to speakers, of 0.5~m, and minimum absolute angle difference of speakers in relation to the listener center axis of at least 20 degrees. All other simulation parameters are kept as before. The training of the neural networks is not re-executed and their parameters are kept exactly the same as for $N=2$ speakers.

As shown in Table~\ref{tab:results}, with $N=3$ speakers, the proposed SSM is robust to changes in the number of speakers and positioning. The NN trained with SSM still outperforms the baselines for almost all cases. When the SNR is low, the NN without SSM again cannot achieve better metrics than the mixed audio obtained at the reference microphone. For high SNR, the baseline NN is able to extract the desired speech given the easier settings. The MVDR filter is clearly affected by the presence of reverberation, as previously observed.

Overall, the results indicate that the use of the proposed speaker selection mechanism in training dramatically improves beamforming results for NN-based beamforming, where the target speaker changes according to the listener's head movement, working well even at a very low SNR level (-10 dB). The SSM is general and robust, providing significant advances toward the solution of the cocktail party problem.

\section{Conclusion}

We proposed a speaker selection mechanism for the training of a neural network model on the task of audio beamforming. The SSM dynamically changes the target speaker in the loss function, at every utterance, focusing on the closest speaker to the listener's head center axis. Through acoustic simulations, the neural network model trained with SSM was able to outperform the baseline NN model (trained without it) and the (ideal) MVDR filter, achieving significantly higher performance metrics. Additionally, we showed that the SSM is robust to changes in the acoustic scene -- number of speakers and positioning. The proposed speaker selection mechanism represents a leap toward the solution of the cocktail party problem. In future work, the employment of the SSM in a real-world set-up is suggested, as well as a deeper analysis of robustness to changes in the acoustic scene.

\bibliographystyle{IEEEtran}
\bibliography{references}

\end{document}